\documentclass[12pt,preprint]{aastex}
\newcommand{\HI}{H~{\sc i}} 
\newcommand{\HII}{H~{\sc ii}}
\newcommand{\kms}{${\rm km~s^{-1}}$}

\newcommand{\VLSR}{\rm{V_{\mathrm{LSR}}}}
\newcommand{\NHI}{\rm{N_{\mathrm{HI}}}}

\slugcomment{To appear in the {\em Astrophysical Journal}}
\shortauthors{LOCKMAN ET AL.} 
\shorttitle{HI and the Nuclear Wind}

\begin{document} 

\title{Tracing the Milky Way Nuclear Wind with 21cm Atomic Hydrogen Emission}

\author{Felix J. Lockman\altaffilmark{1}}
\affil{National Radio Astronomy Observatory, Green Bank, WV 24944}
\email{jlockman@nrao.edu}

\author{N.\ M.\ McClure-Griffiths}
\affil{Research School of Astronomy \& Astrophysics, The Australian National University, Canberra ACT 2611, Australia}
\email{naomi.mcclure-griffiths@anu.edu.au}

\altaffiltext{1}{The National Radio Astronomy Observatory is a facility of the National Science Foundation operated under a cooperative agreement by Associated Universities, Inc.}

%\altaffiltext{1}{National Radio Astronomy Observatory, Green Bank WV
%  24944, USA; jlockman@nrao.edu}
%\altaffiltext{2}{Research School of Astronomy \& Astrophysics, The
 % Australian National University, Canberra ACT 2611, Australia;
 % naomi.mcclure-griffiths@anu.edu.au}

%--------------------------------------------
\begin{abstract}
%-------------------------------------------
There is evidence in 21cm \HI\ emission  for voids several kpc in size 
centered approximately on the Galactic center,
both above and below the Galactic plane.
These appear to map the boundaries of the Galactic nuclear wind. 
An analysis of \HI\ at the tangent points, where the distance 
to the gas can be estimated with reasonable accuracy, shows 
a sharp transition  at   Galactic radii $R\lesssim 2.4$ kpc 
from the  extended neutral gas layer characteristic of much of the Galactic disk, 
to a thin Gaussian layer 
 with FWHM $\sim 125$ pc.  An anti-correlation between \HI\ and $\gamma$-ray
  emission at latitudes $10\arcdeg \leq |b| \leq 20\arcdeg$ suggests that the
 boundary of the extended  \HI\ layer
 marks the walls of the Fermi Bubbles.  With \HI\ we are able to
  trace the  edges of the voids from $|z| > 2$ kpc 
  down to $z\approx0$, where they have a  radius  $\sim 2$ kpc.  
The extended HI layer likely results from star formation in the disk, which 
is limited largely to $R \gtrsim 3$ kpc, so the wind may be
 expanding into an area of relatively little \HI. 
Because the \HI\ kinematics can discriminate between gas in the Galactic 
center and foreground material, 21cm \HI\ emission may be the best probe of the 
extent of the nuclear wind near the Galactic plane.

 \end{abstract}


%----------------------------------------------------------
\section{Introduction}
\label{sec:intro}
%----------------------------------------------------------
The nuclei of galaxies often launch fast
moving winds from active bursts of star formation
or accretion events onto a central supermassive black hole \citep[see the 
review by][]{veilleux05}. The Milky Way appears to be no exception. 
Over the past three decades direct evidence 
for a nuclear wind has been mounting. 
Evidence has come from across the electromagnetic
spectrum, including extended soft X-ray emission
\citep{snowden97,Sofue2000,bland-hawthorn03}, diffuse microwave excess from WMAP
\citep{finkbeiner04,dobler10} and polarized radio continuum
\citep{carretti13}.  The imaging of excess $\gamma$-ray emission
extending to latitudes $|b|\sim 55\arcdeg$ confirmed the existence  
and full extent of the Milky Way's nuclear wind structures, now known
at the Fermi Bubbles \citep{su10,ackermann14}.  Furthermore, there is now direct 
detection of gas within the outflow, measured through absorption lines in ionized species 
towards background AGN \citep{keeney06,Fox15} and in atomic hydrogen emission 
\citep{mcgriff13}.  There is also indirect evidence: 
the \HI\ layer at $R \lesssim 3$ kpc is noticably thin and lacks the 
vertical extensions characteristic of most of the disk \citep{Lockman84} 
suggesting that it might have been excavated by 
 a Galactic wind \citep[e.g.,][]{bregman80}.  

Much about the nuclear wind is unknown including its energy source, which may be an 
outburst from the central black hole, or the product of star formation, either 
episodic or continuous.  Likewise, estimates for the age and timescale for evolution of the wind 
vary by more than an order of magnitude, from a few 10 Myr to a few hundred Myr 
 \citep[e.g.,][]{bland-hawthorn03,su10,Lacki2014,Sarkar2015,Crocker2015}.  
The wind has internal structure and 
is not symmetric, and this manifests itself in the offset of various tracers from each other; there are 
also asymmetries observed that may indicate multiple energetic  events or 
perturbations to the wind structure arising in environmental influences, either 
internal to the Milky Way or external \citep{Crocker2015,Kataoka2015,Sarkar2015}.

It is difficult to discern the shape of the Bubbles at low Galactic latitudes 
using tracers that do not carry kinematic information because of 
confusion with unrelated material and foreground structures such as Loop I.   
 To determine the morphology of the Fermi Bubbles in
$\gamma$-rays, e.g.,  foreground
$\gamma$-ray emission must be estimated and removed.
Here we use the velocity structure of the \HI\ line
together with a model of Galactic kinematics to separate emission originating near 
the Galactic center from that of the intervening medium.  If the Fermi
Bubbles are filled with hot gas, they should appear as voids in 
\HI\ which can be used to map the morphology of the Bubbles at low latitudes.

Throughout this work we adopt the Sun-center distance and LSR rotational velocity 
$R_0 = 8.5$ kpc, and $|V_0| = 220$ \kms, and, for consistency, scale previous work to these 
 values.  Although the earlier results \citep{Lockman84} were obtained using a simple model of 
circular rotation with some density-wave streaming, the subsequent 
characterization of the bar in the Milky Way \citep{Blitz91} allows us to use more 
accurate models for Galactic kinematics \citep[e.g.,][]{weiner99}.

%----------------------------------------------------------
\section{An \HI\  Image of the Galactic Center}
\label{sec:image}
%----------------------------------------------------------
\subsection{The \HI\  data}

We use 21cm \HI\ data from the 
Parkes Galactic All-Sky Survey \citep[GASS;][]{mcgriff09,kalberla10}, 
the most recent version of which has improved spectral baselines \citep{Kalberla2015},  
for investigation of the nuclear wind.  The 
GASS spectra were obtained using the Parkes radio telescope.  The survey has  an angular resolution
of 16\arcmin,   a  velocity channel spacing of $0.8$
\kms, and an rms noise of 57 mK.     The GASS data have uniform 
coverage over the area of this study, at $|\ell| < 30\arcdeg$ and $|b| < 30\arcdeg$ 
for $\delta \lesssim 0\arcdeg$.

\subsection{Terminal Velocities and Tangent Points}

On sight-lines  interior to the solar circle at galactocentric distances ${\rm R < R_0 \equiv 8.5}$ kpc,
the projection of Galactic rotational velocities $V_{\theta}(R)$ 
onto the LSR is greatest at the 
tangent points where $R_t = R_0 \ |\sin(\ell)|$.  These tangent points are at a distance from the Sun 
(at $b \approx 0\arcdeg$)  of 
${d_t = R_0 \ \cos(\ell)}$,  and here any radial or vertical motions, $V_R$ or $V_z$, are projected 
across the line of sight and do not appear in $\VLSR$.  
 Provided that azimuthal streaming motions are not too large, the  maximum measured positive
(negative) velocity of  \HI\ emission in the first (fourth) longitude quadrant, $V_{max}(\ell) $, 
can thus be equated to the terminal velocity from Galactic rotation, $V_t$, and 
the emission assigned to $d_t$ and $R_t$.  
A discussion of potential uncertainties associated with this chain of reasoning 
is given  in the Appendix.

There will always be some \HI\ emission 
at $|\VLSR| > |V_t|$ because of random motions in the interstellar gas, 
and this can be accounted for in determining $V_{max}$.  Another factor 
that must be considered is the  change in $\VLSR$ with distance, which is a strong function of longitude
and distance.  This determines the spatial interval around $d_t$ 
that contributes emission in a velocity range around $V_t$.  
The issues are well understood
\citep[e.g.,][] {BurtonGordon78,Lockman84,Celnik1979,Malhotra1995,mcgriff07}. 
For this investigation we use the \citet{weiner99} model of gas kinematics in the inner Galaxy
(hereafter WS), which includes the dynamical effects of a bar, adjusted to  
 $|V_0| = 220$ \kms\ and $R_0$ = 8.5 kpc.  
At each longitude we evaluate the maximum value of $|\VLSR|$ appropriate to the 
inner Galaxy and set that equal to $|V_t|$. 
  
Our general results do not depend on specific properties of the WS bar model; use of 
the models of \citet{Fux99} or \citet{Rodriguez-Fernandez08} would lead to 
qualitatively identical conclusions.   The results 
are also robust to assumptions about the exact Galactic rotation curve, 
the existence of density-wave streaming motions, or Galactic constants.

\section{Voids in \HI\ Above and Below the Galactic Plane}

\begin{figure}
\centering
\includegraphics[width=2.5in]{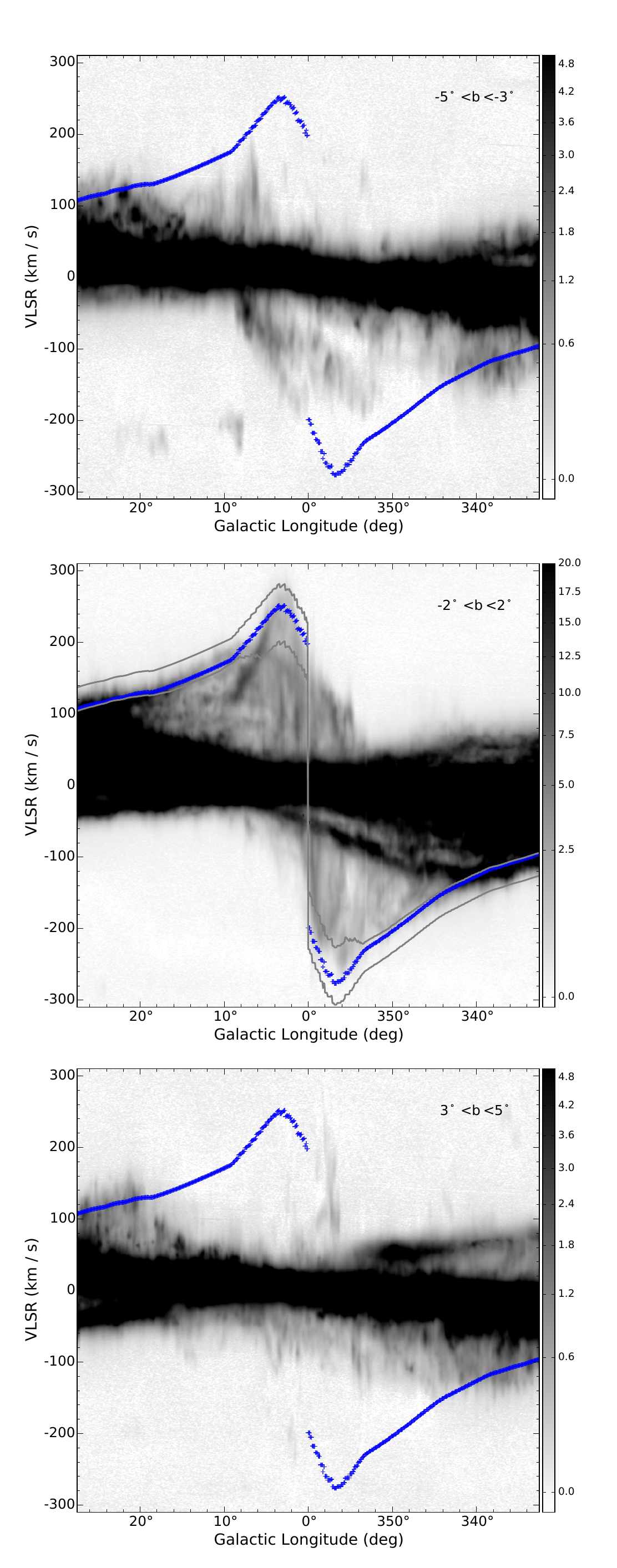}
\caption[]{Longitude-velocity images of the averaged \HI\ brightness temperature in the
  ranges $3\arcdeg<b<5\arcdeg$ (top), $|b|<2\arcdeg$ (middle), and
  $-5\arcdeg<b<-3\arcdeg$ (bottom), overlaid with the terminal velocity values
 from the WS model  (blue crosses) used to identify tangent point velocities.  The \HI\ is displayed with a
  square root transfer function in Kelvin, as
  shown in the wedges to the right.  The grey lines
around $V_t$ in the central panel show the velocity
  limits used to integrate the emission included in  Figure~\ref{fig:image}, and apply 
at all latitudes.   At $b\approx0$ these velocity intervals encompass \HI\ emission at
all longitudes.  However,  there is little \HI\ emission above and below the plane 
around $V_t$ in the inner $\pm15\arcdeg$ of longitude.  The absence of emission 
over a considerable range in $\VLSR$ indicates that the  voids are quite large.  At the tangent 
points the average distance from the Galactic plane in the upper and lower panels varies from 0.4 kpc 
to 0.5 kpc.
\label{fig:lv}}
\end{figure}

Figure~\ref{fig:lv}  shows the \HI\ emission as a function of longitude and $\VLSR$,
with values of $V_{max}$ from the WS model as blue points.  
%The grey lines define the velocity range chosen for the tangent-point analysis.  
The central panel shows \HI\ 
averaged over latitudes $|b|<2\arcdeg$, while the upper and lower panels show the same 
curves on \HI\  emission averaged over $3\arcdeg < b < 5\arcdeg$ and $-5\arcdeg < b < -3\arcdeg$,
respectively.  

It is apparent that at $b \approx 0\arcdeg$  (central panel) velocities around $V_{max}$ contain \HI\ emission 
at all longitudes, but only a few degrees above and below the plane there are 
large voids in $(\ell-V)$ space  spanning more than $15\arcdeg \times 100$ \kms. 
Within these voids there is only occasional \HI\ emission; it can be identified with non-circular motions 
arising from the Galactic bar or conditions in the inner nucleus 
\citep[e.g.,][]{Oort77,BurtonLiszt78,Binney91,Rodriguez-Fernandez08}.

It is known that extra-planar gas in many spiral galaxies has a vertical lag in rotational velocity 
${\rm dV_{\theta}/dz \approx -10}$ \kms\ kpc$^{-1}$ \citep{Sancisi01}.  The lag 
likely originates from the interaction between \HI\ clouds and a more
slowly rotating hot galactic corona \citep[e.g.,][]{Fraternali2008, Melioli2009,Marasco2015}. 
In the Milky Way a measurement about 3 kpc above a superbubble gives a lag of -8 \kms\ kpc$^{-1}$ 
while model fits to large-scale \HI\ surveys find a value about twice as large, -15 \kms\ kpc$^{-1}$ 
\citep{Pidopryhora2007,Marasco2011}.     
  A lag of this magnitude 
cannot account for the voids in Fig.~\ref{fig:lv}, which appear abruptly at $\ell \sim\pm15\arcdeg$ 
from the Galactic center and are fully formed by $|z| \approx 0.25$ kpc.  
To create the voids from a rotational lag, 
 Galactic rotation would have to be reduced by more than 100 \kms\ at $|z| = 0.25$ kpc, 
and would have to fall   to near zero at $|z| \approx 0.5$ kpc.  
 In actuality, ample gas with a very small lag, if any, is found at $z = 0.5$ kpc 
 for $R \gtrsim 3$ kpc \citep[e.g.,][]{Lockman2002,Ford2010}.  

The voids in longitude and velocity  thus imply that there are  large volumes of space with little neutral 
 Hydrogen in the inner Galaxy.  
  To quantify the extent of the voids at all locations in the inner Galaxy 
would require a detailed model of the Galaxy's kinematics and the result would not be unique, 
but the basic parameters of the voids can 
be determined from an analysis along the tangent points, 
as identified by  $V_{max}$, where the distance can be determined with 
reasonable accuracy.

Setting $V_t = V_{max}$ derived from the WS model,  
at each longitude and latitude we
have calculated  the tangent-point \HI\ emission over a velocity interval around $V_t$. 
 In the first quadrant of longitude this can be written
\begin{equation}
  N_H(\ell,b) = 1.823 \times 10^{18}\, \int_{V_t-\Delta v_1}^{V_t+\Delta v_2}
  \tau(v)\, T_b(v)\, dv \,~{\rm cm^{-2}}.
\label{eq:NH}
\end{equation}
This equation was also applied to the fourth longitude quadrant with a suitable change of sign in the velocities. 
The velocity interval $\Delta v_1$ is chosen from the WS model as described in
the Appendix to encompass a line-of-sight distance of approximately $1$
kpc around the tangent point.  It takes a value of $3$ \kms\ for
longitudes $|\ell|>20\arcdeg$, increases slowly between $|\ell| = 20\arcdeg$
and $|\ell| = 8\arcdeg$ to $ 10$ \kms, increasing rapidly thereafter
to $50$ \kms\ at $|\ell| = 5\arcdeg$.  Because of the substantial non-circular 
motions that are associated with the Galactic nucleus near 
$\ell=0\arcdeg$, we arbitrarily limit $\Delta v_1$ to $50$ \kms\ within 
$5\arcdeg$ of the Galactic center, with consequences discussed in the Appendix. 
The innermost part of the Galactic disk is associated with large non-circular 
motions in all species \citep[e.g.,][]{Burton93}, so 
our results lose  accuracy at $R \lesssim 1$ kpc.
The upper velocity limit,
$\Delta v_2=30$ \kms, is held constant with longitude and was chosen to capture 
most of the \HI\ emission in the wings of the profiles \citep{kulkarni85,mcgriff07}.
The velocity limits 
are shown as grey lines around $V_t$ in the central panel of Figure~\ref{fig:lv}.  
We estimate the optical depth per channel from the brightness temperature, $T_b$,
of the emission, where $\tau(v) = -\ln{(1 - T_b(v)/T_s)}$, with an
assumed gas spin temperature $T_s=150$ K \citep[e.g.,][]{DickeyLockman90}.
Figure~\ref{fig:image} shows the optical depth
corrected column density  as a function of longitude and latitude.

Figure~\ref{fig:image} shows  that there are two large cavities 
in Galactic \HI\ at $|b|\gtrsim 2\arcdeg$ over longitudes 
$-15\arcdeg \lesssim \ell \lesssim 17\arcdeg$.  Within these cavities
there is almost no detectable \HI\ emission with the exception of a
few small, isolated clouds that have typical column
densities of $\NHI<7\times 10^{18}~{\rm cm^{-2}}$, likely related to the population of \HI\ clouds 
discovered to be entrained in the
nuclear wind \citep{mcgriff13}. For the adopted Sun-center distance 
$R_0 = 8.5$ kpc,  the region devoid of \HI\ extends from
$R\approx 2.1$ kpc on the negative longitude (Southern) side of the
Galactic center to $R\approx 2.4$ kpc on the positive longitude
(Northern) side of the Galactic center.  

The general features of this Figure do not depend on the details of the adopted $V_t(\ell)$ 
function or the precise values of $\Delta v_1$ and $\Delta v_2$.  As is obvious from Fig.~\ref{fig:lv},
the large voids above and below the plane will map into spatial voids centered approximately  on the Galactic 
center.  

\subsection{Gas Distribution and Mass}
Figures \ref{fig:lv} and \ref{fig:image} confirm the  absence of gas in an extended
\HI\ layer  within 3 kpc of the Galactic center.
To demonstrate this more quantitatively, we plot the distribution of mean
column density, $\left<\NHI \right>$, versus distance from the Galactic plane, $z$, for
several ranges of galactocentric radius in Figure~\ref{fig:HI_z}.  
The column density is from
Figure~\ref{fig:image} in intervals of R averaged over both positive and
negative Galactic longitudes.  The distribution within $R<2$ kpc is
well described by a single Gaussian of FWHM $\sim 125$ pc, whereas at
$R>2.75$ kpc the disk and lower halo require the well-known two
Gaussian components with FWHM $\sim 150$ pc and $\sim 400$ pc, plus an
extended exponential tail of $\sim 500$ pc 
\citep{Lockman84, Lockman86,SavageMassa87,DickeyLockman90,SavageWakker09}. 
 In the transition region around $R\sim 2$--$2.5$
kpc the distribution is described by two Gaussians of width 100 and 300 pc, without an additional exponential
component.  These functions are drawn in Fig.~\ref{fig:HI_z} as dashed lines.

To convert from a column density measured in the plane of the sky to
either a number density of hydrogen or a surface density
% $\Sigma_{HI}$ integrated through $z$,
 it is necessary  to know the distance,
$\Delta d$, that contributes to the emission at the terminal velocity.
As described above,  the velocity interval for the
column density calculation was chosen to give a depth $\Delta d \approx 1$ kpc
at all longitudes.  We use this to estimate the average number density
of \HI, $n(z)$, in bins of radius.  The surface density,
$\Sigma_{HI}$, is then $n(z)$ integrated through the $z$ direction for each
radius bin.  These values are given in Table~\ref{tab:surf_dens}.
 Because of the uncertainties in converting from a velocity interval to $\Delta d$, the
errors on the surface density may be as large as a factor of two,
but even so, it is clear that there is a significant decrease in \HI\ 
surface density towards the center of  the Milky Way, 
nearly an order of magnitude decrease between $R>3$ kpc and $R<2$ kpc.

\section{Comparison with the Fermi Bubbles}
Overlaid on the \HI\ image of Fig.~\ref{fig:image}  is  the
\citet{su10} template for the Fermi Bubble edges (red circles) and their
so-called ``northern arc'' (cyan circles), which is an excess of $\gamma$-ray
emission surrounding the positive longitude edge of the Bubble.  
For $\ell < 0\arcdeg$ and 
$|b|>10\arcdeg$  the \HI\ void edges match the \citet{su10} Fermi Bubble
 templates reasonably well.  In contrast, 
at $\ell > 0\arcdeg$  the \HI\ fails to match the Bubble template at any 
latitude; the Fermi Bubble template lies well within the 
\HI\ void.  However, the \HI\ emission at positive longitudes and $b > 0\arcdeg$ does
match  the inner edge of the so-called ``Northern Arc''.
Because of uncertainties in modelling and subtracting the
foreground, the \citet{su10} template at $|b|<10\arcdeg$ is necessarily uncertain, 
and  more recent work on the morphology of the Fermi Bubbles by
\citet{ackermann14} does not attempt to quantify the Bubble
emission within 10\arcdeg\ of the Galactic plane.

If the Fermi Bubbles are filled with a hot plasma the absence of \HI\
within the Bubble structures is not surprising.  
%% To address the correspondence we compare the \HI\ and $\gamma$-ray emission.
Figure~\ref{fig:HI_Fermi} is a comparison between the \HI\ emission
averaged in latitude bins and the Fermi $\gamma$-ray emission
\citep{ackermann14} in the same bins.  
The \HI\ data are the same as shown in Fig.~\ref{fig:image}.  Systematic errors in the 21cm 
spectral baselines produce systematic errors in the derived ${\rm \langle N_{\rm HI} \rangle}$ at the 
level of 10$^{18}$ cm$^{-2}$. This corresponds to a baseline offset of $<1\sigma$ of the channel 
noise over 20 \kms. There is clearly an anti-correlation  between the $\gamma$-ray 
emission and the \HI\ emission:
the \HI\ emission is absent throughout the longitude range that shows GeV
emission but appears  where the Fermi $\gamma$-ray emission
decreases.

Uncertainties in the foreground subtraction required for extracting
$\gamma$-ray emission make estimates of the shape of the Fermi
Bubbles at $|b|<10\arcdeg$ from high energy observations prone to significant errors.  The \HI\ kinematics,
on the other hand, allows a clear separation of the emission originating 
at small Galactic radii from emission which is unrelated.  
 Moreover, the $\gamma$-ray emission that traces the Fermi Bubbles does not necessarily 
give a complete picture of the location and extent of the nuclear wind volume, 
as shown in recent models for the Bubbles \citep{Crocker2015,Sarkar2015}.
At low latitudes the \HI\ distribution should be a better tracer of the
shape of the nuclear wind region than the $\gamma$-rays and the Fermi Bubbles.

\section{Comparison with Star Forming Regions}

The extended \HI\ layer so visible in Fig.~\ref{fig:image} at $|\ell| \gtrsim 15\arcdeg$ almost
certainly results from large-scale star formation processess, as verified by 
correlations observed both in the Milky Way and a variety of other galaxies  
\citep{Ford2010,Heald2015}, though we lack a detailed understanding of the process.  
As the rate of star formation varies 
significantly across the Galaxy, it is worthwhile to consider whether the voids we 
describe might result in some part from the absence of 
an extended \HI\ layer owing to a reduced star formation rate in the inner Galaxy.

We take as the measure of recent star formation the location of \HII\ regions 
in the inner Galaxy from the WISE catalog \citep{Anderson2014}.  
In Fig.~\ref{fig:image}  the yellow dots show the longitude and 
latitude of \HII\ regions selected under the same kinematic criteria 
used to select the \HI. The \HII\ regions
 are prevalent at the tangent points, and wherever there are \HII\ regions 
there is the extended \HI\ layer.  The extended \HI\ layer, however, does not cut 
off at the edge of the star forming regions but extends to lower $|\ell|$, 
and thus closer to the Galactic center.

Fig.~\ref{fig:HII_vs_R} shows the surface density of  \HII\ regions 
over  $6\fdg2 \leq |\ell| \leq 80\arcdeg$ 
against distance from the Galactic center, derived 
assuming a flat rotation curve with V$_{\theta} = 220$ \kms.  The vertical line shows 
the approximate boundary of the extended \HI\ layer.  Although there is a general correlation  
between the absence of \HII\ regions at small values of R and the absence of an extended \HI\ 
layer, it again appears that the \HI\ is found much closer to the Galactic center
 than the \HII\ regions.   The \HI\ voids do not seem to be simply a consequence of the low star 
formation activity at $R\lesssim 2.4$ kpc.  Moreover, the boundary of the \HI\ void is relatively sharp 
(Fig.~\ref{fig:HI_Fermi}) whereas the \HII\ region surface density decreases gradually at $R<4$ kpc.  
 Still, the absence of significant star formation at $0.2 \lesssim R \lesssim 3.5$ kpc suggests that the 
nuclear wind may be expanding though a medium that initially had less of an extended \HI\ component than 
exists throughout the main Galactic disk.

\section{Discussion and Summary}
The absence of extra-planar \HI\ within $R\sim 2.6$ kpc of the Galactic center
was first noted by \citet{Lockman84}. 
Using more modern data, we find that at 
$R\lesssim 2.1$ kpc on the Southern side of the Galactic center and
$R\lesssim 2.4$ kpc on the Northern side the \HI\ layer is well described
by a single Gaussian component of FWHM $\sim 125$ pc.  Beyond $R> 3.5$
kpc the distribution returns to the well-known double Gaussian (FWHMs
$\sim 150$ pc and $400$ pc) plus an exponential tail.  At 
intermediate radii the distribution is better described by  the
double Gaussian alone.  The total surface density of \HI\ is 
nearly an order of magnitude  smaller at $R<2$ kpc than it is at 
$R\geq3.5$ kpc.  A decrease in $\Sigma_{HI}$ is often
observed in the inner regions of spiral galaxies including the Milky Way
\citep[e.g.][]{Roberts67,wong02,BurtonGordon78}, 
but here we have the favorable viewing angle that allows us to show that for the Milky
Way about 20\% of the missing \HI\ can be attributed to the absence of the vertically
extended \HI\ layer, which terminates at a relatively sharp boundary (Fig.~\ref{fig:HI_Fermi}).

These results are based on an analysis of \HI\ kinematics, and thus 
rely on the assumption that the WS model (or any model that 
matches the terminal velocities in the plane, \citep[e.g.,][]{Fux99,
Rodriguez-Fernandez08}) applies not only in the plane 
but vertically for several hundred parsecs.    The \HI\ data alone cannot rule out 
the existence of an extreme vertical kinematic anomaly such as a rotational lag of 
$ -400$ \kms\ kpc$^{-1}$ in the inner Galaxy.  However, the general symmetry of the kinematic 
voids, their rapid onset $\sim 15\arcdeg$ from the center, their correlation with 
the edges of the Fermi Bubbles, and the detection of \HI\ and ionized gas 
apparently entrained in the nuclear wind \citep{mcgriff13,Fox15}, all  
suggest that the simpler picture of a spatial void 
in extra-planar \HI\ around the Galactic center is more likely.

A high energy, hot, nuclear wind should excavate diffuse  \HI\ 
from its vicinity and thus \HI\ emission should be
anti-correlated with the hot $\gamma$-ray emitting plasma 
\citep{veilleux05}.   At Galactic latitudes
$|b|>10\arcdeg$, where the excess $\gamma$-ray emission of the Fermi
Bubbles has been carefully extracted, we have found that the \HI\ is very well
anti-correlated with the $\gamma$-ray emission.  Because the kinematics of \HI\ allows
spatial separation of Galactic center emission from unrelated material, it should provide the
best tracer of the extent of the Fermi Bubbles,
 and hence the nuclear wind,  within a few degrees
of the Galactic plane, at the latitudes where foreground modelling
makes estimating the high-energy emission particularly difficult.

Because the \HI\ voids occur in the inner Galaxy where, outside of the innermost few 
hundred pc,  the rate of star formation is low, the nuclear wind likely expanded  
through an ISM considerably less dense and extensive than is found further from the 
Galactic center.

The images we have provided give an indication of the somewhat
surprising morphology of the nuclear wind, which, if we 
interpret the dimensions of the \HI\ voids as deriving entirely from the wind, 
 appear to extend to
$R\sim 2.4$ kpc at $z\approx0$.  Although the implied opening angle
 close to the plane may seem large, a recent hydrodynamic model of
the Fermi Bubbles shows them extending to almost $R\sim 2$ kpc at
$z\sim 1$ kpc \citep{Sarkar2015}.  
%To assist future models of the evolution of the Bubbles we have included estimates of \HI\ column density.
Whatever the source of the Fermi Bubbles, it has apparently left its mark on the large-scale distribution of
 \HI\ in the inner Galaxy.

\acknowledgements The Parkes Radio Telescope is part of the
Australia Telescope which is funded by the Commonwealth of Australia
for operation as a National Facility managed by CSIRO.  We are
grateful to D.\ Malayshev and A.\ Franckowiak for providing the Fermi GeV
data in electronic form.  We thank  R.\ Benjamin, R.\ Crocker, 
C.\ Federrath, and H.A. Ford for valuable conversations while preparing this work.

{\it Facility:} \facility{Parkes}

\appendix
\section{$V_{max}$, the Terminal Velocity, and the Tangent Point}

Using a right-handed cylindrical coordinate system with the Sun at $x = -R_0$, the velocity of an object 
with respect to the LSR is 

\begin{equation}
V_{LSR} = \left[R_0 \sin{\ell} \left(\frac{V_0}{R_0}-\frac{V_{\theta}}{R}\right) + V_R
\cos{(\ell+\theta)}\right]\, \cos{b} + V_z \sin{b},
\label{eq:VLSR}
\end{equation}
where  $R_0$ is the Sun-center distance, taken to be 8.5 kpc, and $V_0$ the 
circular velocity of the LSR, taken  to be $-220$ \kms\  in this coordinate system.  
The azimuthal coordinate $\theta$ runs 
opposite to Galactic rotation and thus $\theta \approx \ell$ for an object at a great distance.   
Velocity components are $V_R$, $V_{\theta}$, and $V_z$. 
At a tangent point the distance from the Sun is $d_{t} = R_0  \ cos(\ell)$.  Here also 
 $R_{t} = R_0 |\sin{\ell}|$,   and $ \theta - \ell = 90\arcdeg$, so 
close to the Galactic plane where $\cos{b} \approx 1.0$ and $\sin{b} \approx 0$,
 eq.~\ref{eq:VLSR} reduces to 
\begin{equation}
V_{LSR} = \left[-V_{\theta} R_0/R_{t} + V_0\right] \sin(\ell)\ \equiv \ V_t.
\end{equation}

As $R_t$ is the  minimum distance from the Galactic center  at all longitudes $|\ell|<90\arcdeg$,
unless V$_{\theta}$ has a large variation over a fairly small distance,  V$_{LSR}$   will  have its 
largest  value at R$_t$, the tangent point: a positive velocity in the first quadrant of longitude, and a negative 
velocity in the fourth.    It is therefore straightforward to identify emission associated 
with a particular tangent point and thus at a given distance.

In the Milky Way, however, streaming motions caused by density waves or the 
bar may produce conditions where the maximum velocity 
does not occur at the tangent point, but at another location,
effecting the apparent rotation curve \citep[e.g.,][]{Burton93}.  This issue was considered recently by 
\citet{Chemin2015} who use the simulation of Renaud et al. (2013) to conclude that although streaming 
may produce errors in the derived rotation curve, for $R\leq6$ kpc the difference between the location of 
emission at $V_{max}$ and the tangent point is only a few percent.  
As we are interested in a more general situation of associating a particular volume 
of space with the terminal velocity we elaborate on this issue for the 
bar model of \citet{weiner99} (WS) used in our analysis.

We generated a set of test particles with a uniform surface density at $R<R_0$ that follow 
 the WS kinematics and have a random velocity component 
 $\sigma_{cc}$ as well.  We 
calculate the particles $\ell, b, z,$ and $\VLSR$, then analyze their properties in a manner identical to the way 
the \HI\ data were analyzed.  This allows us to determine how well our selection of tangent point gas
 represents the actual sitation at that location.

Figure~\ref{fig:WS_lv}  shows the results of one of the simulations for a velocity dispersion
 $\sigma_{cc} = 7$ \kms\  in R, $\theta$, and z.
   Particles  located within $\pm0.6$ kpc of 
the tangent points are colored red.  Virtually all of the red points are located along the velocity limits of 
the simulation showing that, execpt for quite low longitudes, the tangent point sample occurs 
near $V_{max}$.  Analyzing the simulation, we find that the location of $V_{max}$ differs from the
 tangent point by usually $\sim 5\%$  though at a few positions near $\ell \sim 20\arcdeg$
 the error can be as high as 20\%.  
This result  is in general agreement with \citet{Chemin2015}.   Errors in the kinematic distance 
to individual objects located some distance from the tangent point can, however, be large, as illustrated 
by the HII region W31 \citep{Sanna2014}.

An equally important consideration for our work is the relationship between a velocity 
interval around $V_{max}$ and the associated distance along the line of sight
 over which these velocities arise, parameterized by the quantity $\Delta v_1$ 
in eq.~\ref{eq:NH}.   The importance of this in analyzing Galactic \HI\ 
 was emphasized by \citet{Burton72}.  While knowledge of $\Delta v_1$  is needed 
to convert column densities to volume densities, for this investigation 
it is more important  that we do not artifically produce structures (or remove them) 
by grossly changing the sample volume with longitude or R.  Evaluating the WS model 
over the longitude range of interest, a 1 kpc  distance corresponds to 
values for $\Delta v_1$ that increase approximately linearly from  
3 to 10 \kms\ between $\ell = 20\arcdeg$ and $\ell = 8\arcdeg$, and 
between 10 and  50 \kms\ from 
$\ell = 8\arcdeg$ to $\ell = 5\arcdeg$.  For $\ell \geq 20\arcdeg$ we use a constant value of 
$\Delta v_1 = 3$ \kms\  which increases the distance along the line of sight over which 
we average emission by a factor $\approx 1.5$.  At $\ell \leq 5\arcdeg$ a constant value 
$\Delta v_1 = 50$ \kms is adopted.  

\begin{figure}
\centering
\includegraphics[width=6in]{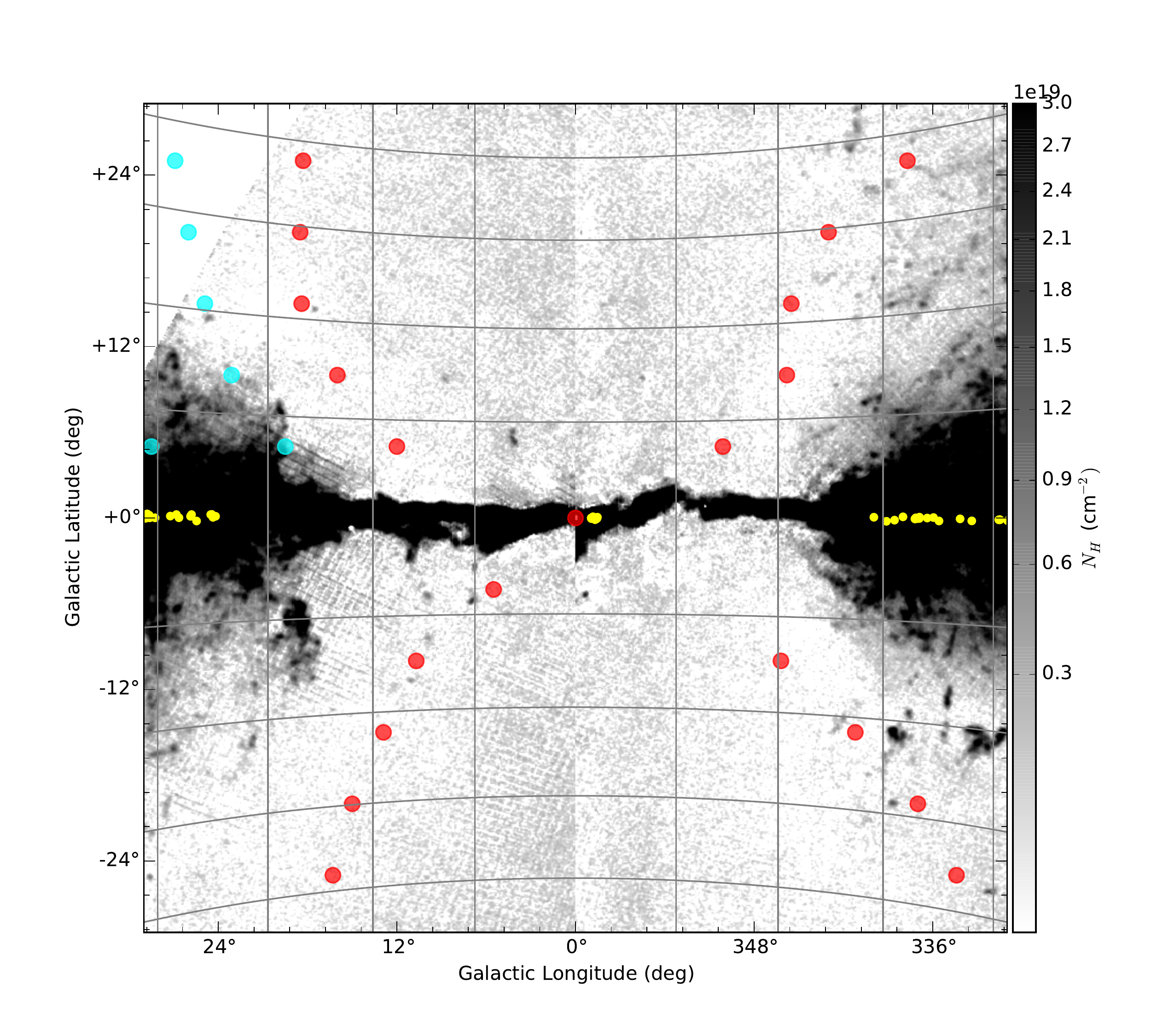}
\caption[]{\HI\ column density along the Galactic tangent points
 showing the absence of a vertically 
extended \HI\ layer in the region around the Galactic center.
 The grid marks 1 kpc 
intervals in distance from the Galactic center and distance from the plane.  
The red dots show the outline   of the Fermi Bubble template and the cyan dots show the outline of
  the ``Northern Arc'', both as determined by \citet{su10}.
 The  greyscale is displayed in the wedge at the right
  with a square root transfer function.  \HI\ column
  densities in the range $|l|<5\arcdeg$ ($R \leq 0.75$ kpc) are not reliable because of
 large non-circular motions associated with the Galactic center.  The yellow dots mark the 
locations of tangent-point \HII\ regions from the WISE catalog \citep{Anderson2014} selected using the same 
kinematic filter used for the \HI\ (eq.~\ref{eq:NH}).

  \label{fig:image}}
\end{figure}

\begin{figure}
\centering
\includegraphics[width=6in]{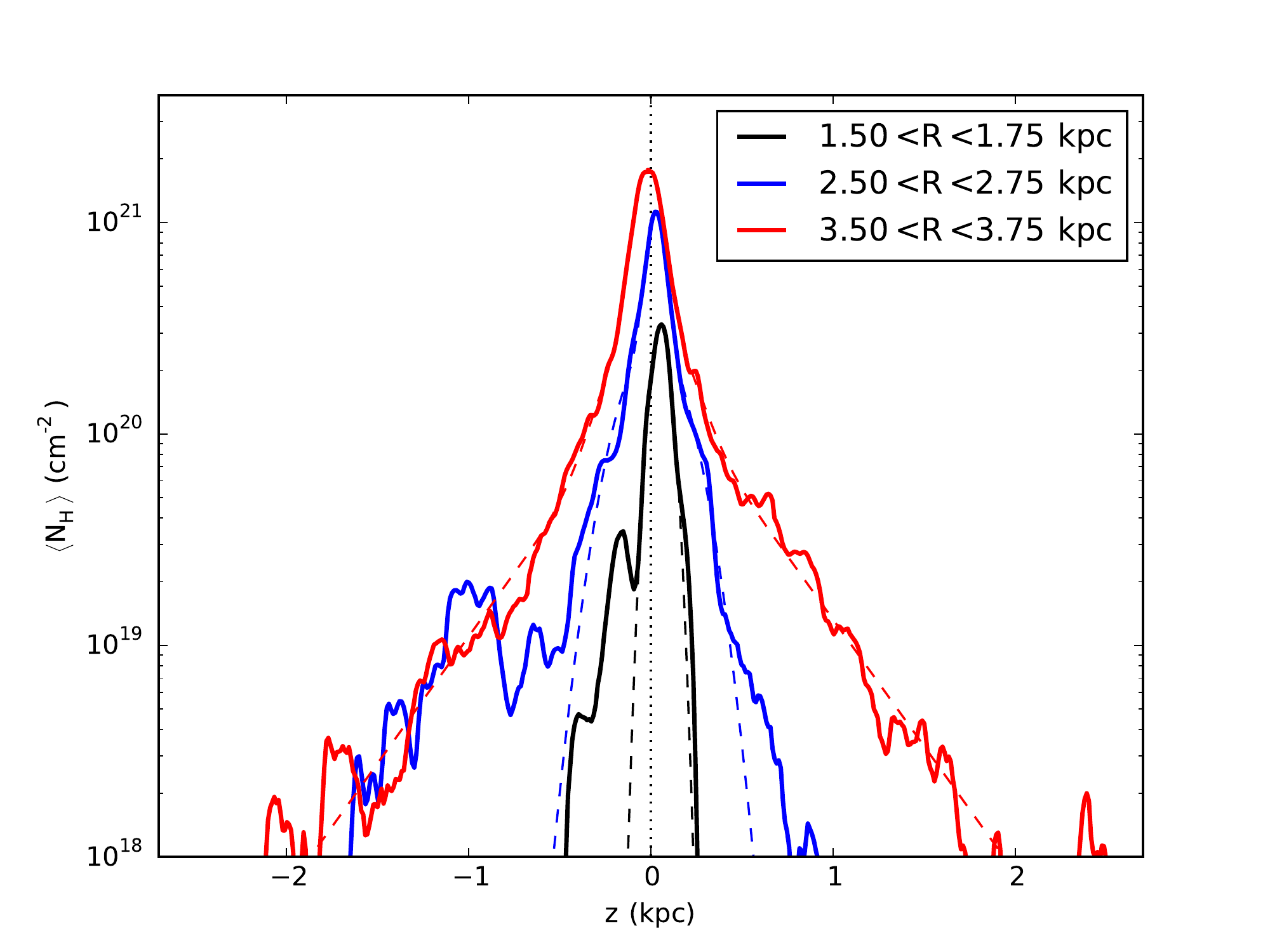}
\caption[]{Average \HI\ column density vs. distance from the Galactic plane in three bins of galactocentric
  radius, R, showing the absence of the extended \HI\ layer towards the Galactic center. 
For $R<1.75$ kpc (black line) the \HI\ distribution is  described by a single
  Gaussian component of FWHM $\sim 125$ pc.  At
  $R >3.5$ kpc (red line), \HI\ distribution is described by
  the well-known two Gaussian components (FWHM $\sim 150$ pc and $\sim
  400$ pc) plus an extended exponential. At the intermediate radii
  (blue line) the distribution is reasonably well-described by two Gaussian
  components, but without the exponential.  These curves include data from both Northern and Southern longitudes. 
Dashed lines show the values of the model fits.
  \label{fig:HI_z}}
\end{figure}

\begin{figure}
\centering
\includegraphics[width=4.5in]{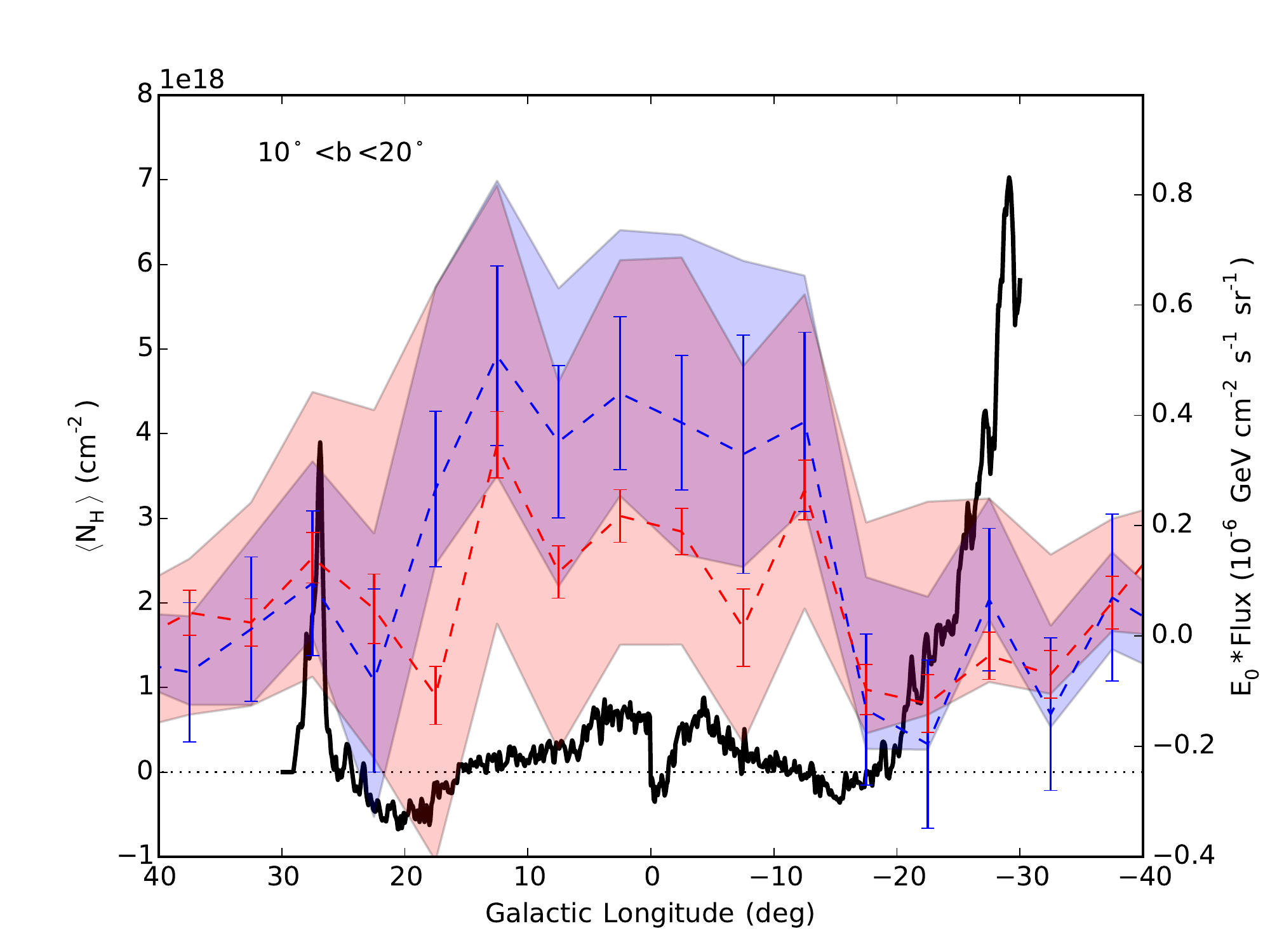}
\includegraphics[width=4.5in]{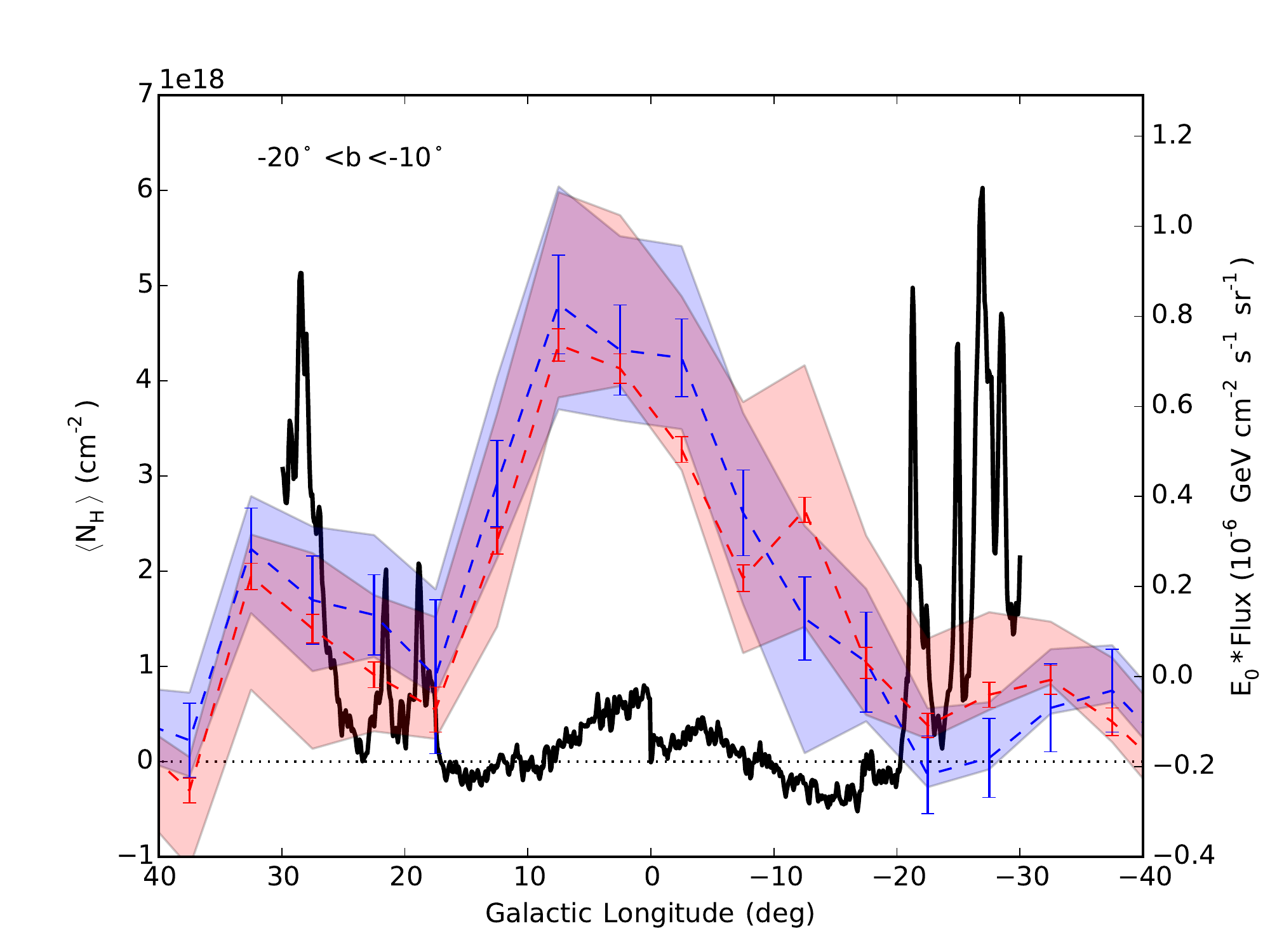}
\caption[]{ Comparison of the mean \HI\ emission with Fermi LAT
  $\gamma$-ray emission at energies $3<E<10$ GeV (red) and $E>10$ GeV
  (blue) reproduced from   \citet{ackermann14} 
  in two latitude bins  $10\arcdeg < b < 20\arcdeg$
  (top) and $-20\arcdeg < b < -10\arcdeg$ (bottom).  The Fermi
  data are the Fermi Bubble residuals after subtraction of the
  GALPROP model and the shaded regions correspond to different
  foreground models as described in \citet{ackermann14}.  \HI\ column
  densities in the range $|l|<5\arcdeg$ are not reliable because of
  unmodelled kinematic effects associated with the Galactic center. 
The anti-correlation between \HI\ and  $\gamma$-ray emission suggests that 
the \HI\ voids trace the Fermi Bubbles.  The figures average over distances from 
the plane of $1.5 \lesssim |z| \lesssim 3.0 $ kpc. 
 Systematic baseline errors in the \HI\ spectra below the 
$1\sigma$ channel noise produce offsets in $\langle N_{HI} \rangle$  
with a magnitude $\sim 10^{18}$ cm$^{-2}$.

\label{fig:HI_Fermi}}
\end{figure}

\begin{figure}
\centering
\includegraphics[width=5in]{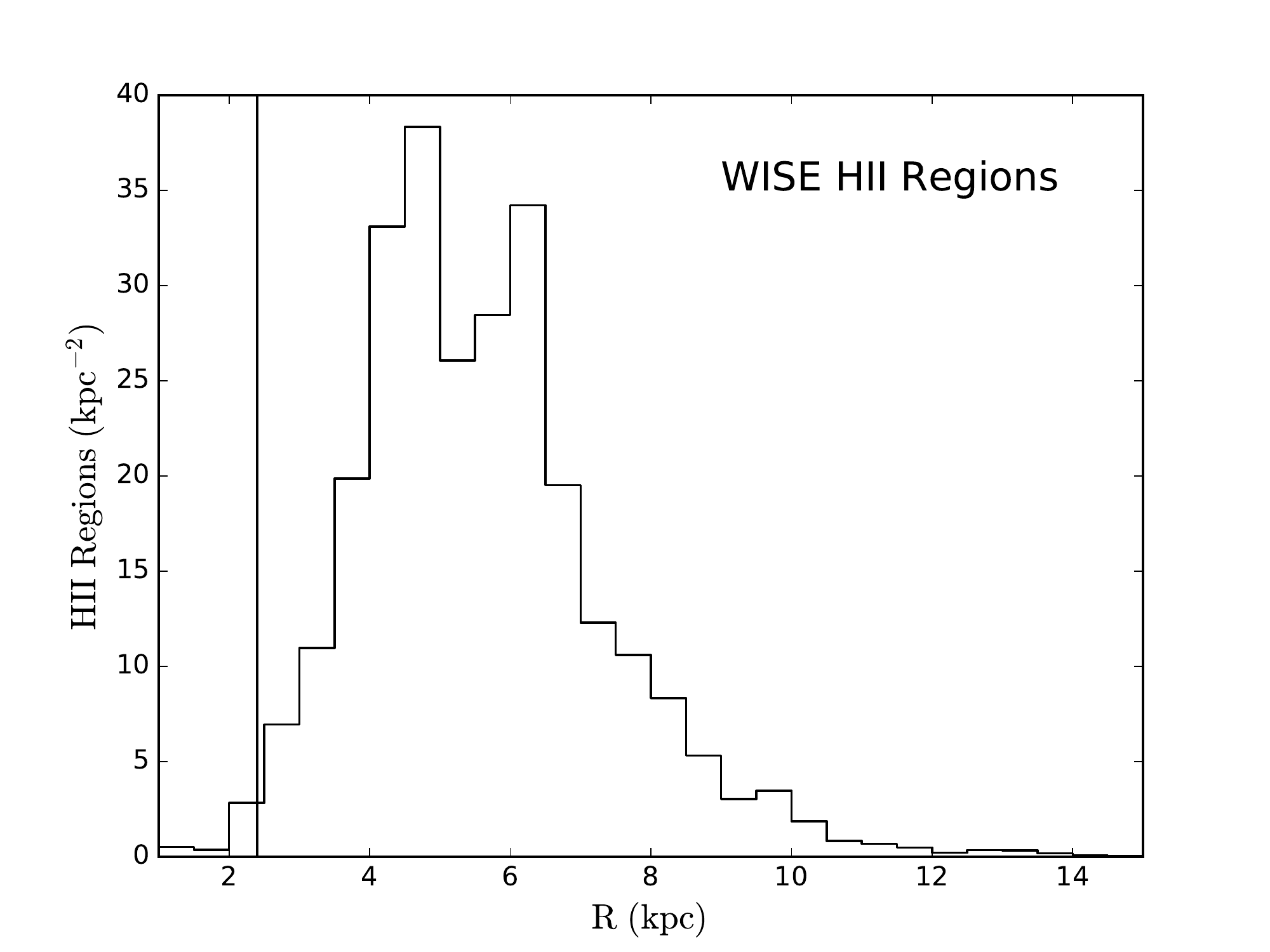}
\caption[]{Surface density of \HII\ regions from the WISE survey 
\citep{Anderson2014} vs. distance from the Galactic center, R, analyzed over $6\fdg2 \leq |\ell| \leq 80\arcdeg$ 
for a flat rotation curve with V$_{\theta}$ = 220 \kms.  The vertical line at R = 2.6 kpc marks the approximate 
edge of the extended \HI\ layer, which is found to smaller values of R than the \HII\ regions.  
\label{fig:HII_vs_R}}
\end{figure}

\begin{figure}
\centering
\includegraphics[width=5in]{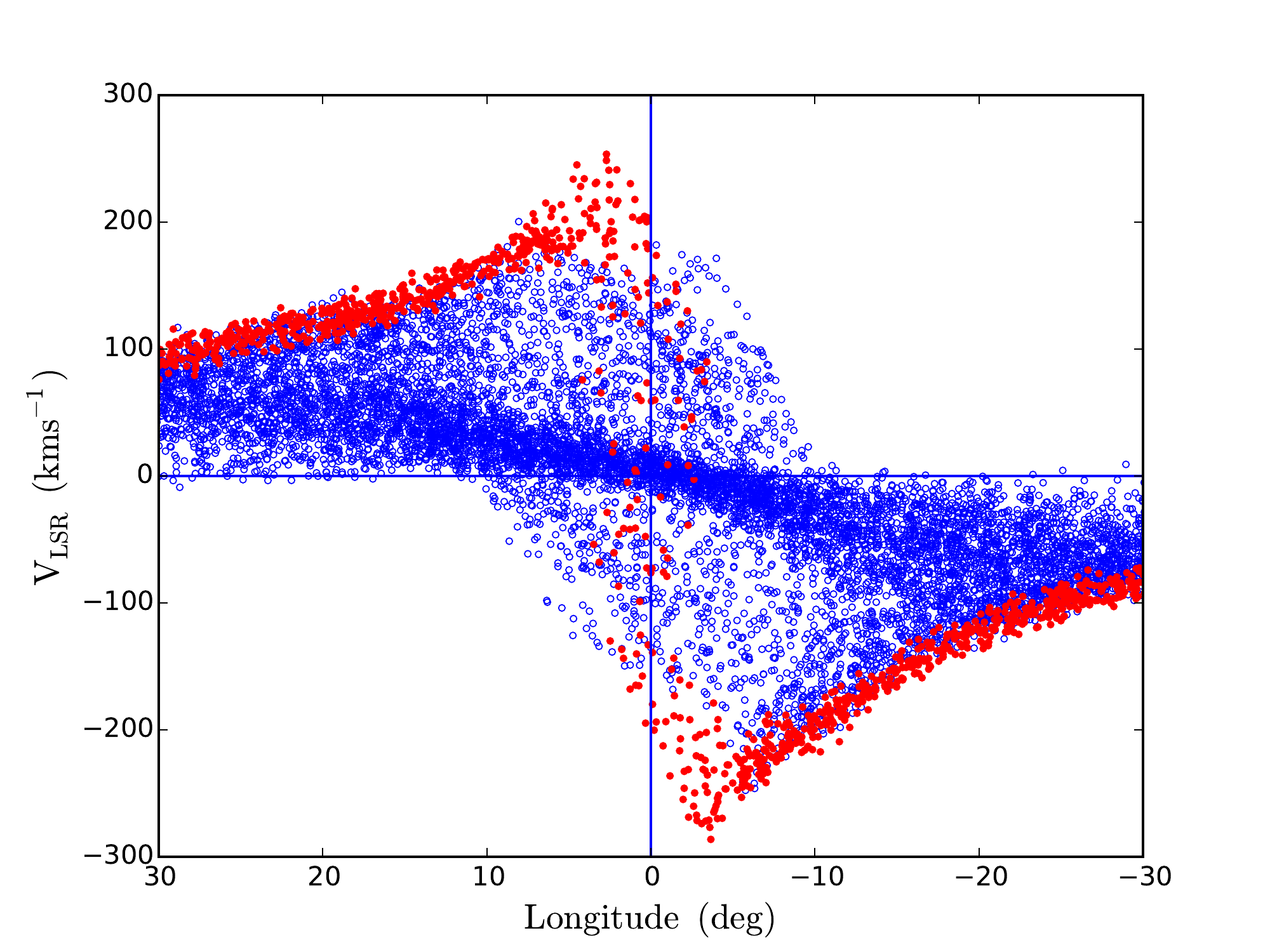}
\caption[]{Velocity-longitude distribution of particles in 
a simulation using the WS kinematical bar model.  
Clouds within $\pm0.6$ kpc of the tangent point are marked in red.  Except quite near 
the Galactic center, the tangent-point clouds custer around the maximum velocity in their 
direction. Selection of \HI\ at the maximum velocity, therefore,  gives a sample of gas near the 
tangent points.  To reduce confusion in this Figure near $\VLSR \approx 0$, only a random subset of the particles at 
$R > 5$ kpc are plotted.
\label{fig:WS_lv}}
\end{figure}

\begin{figure}
\centering
\includegraphics[width=5in]{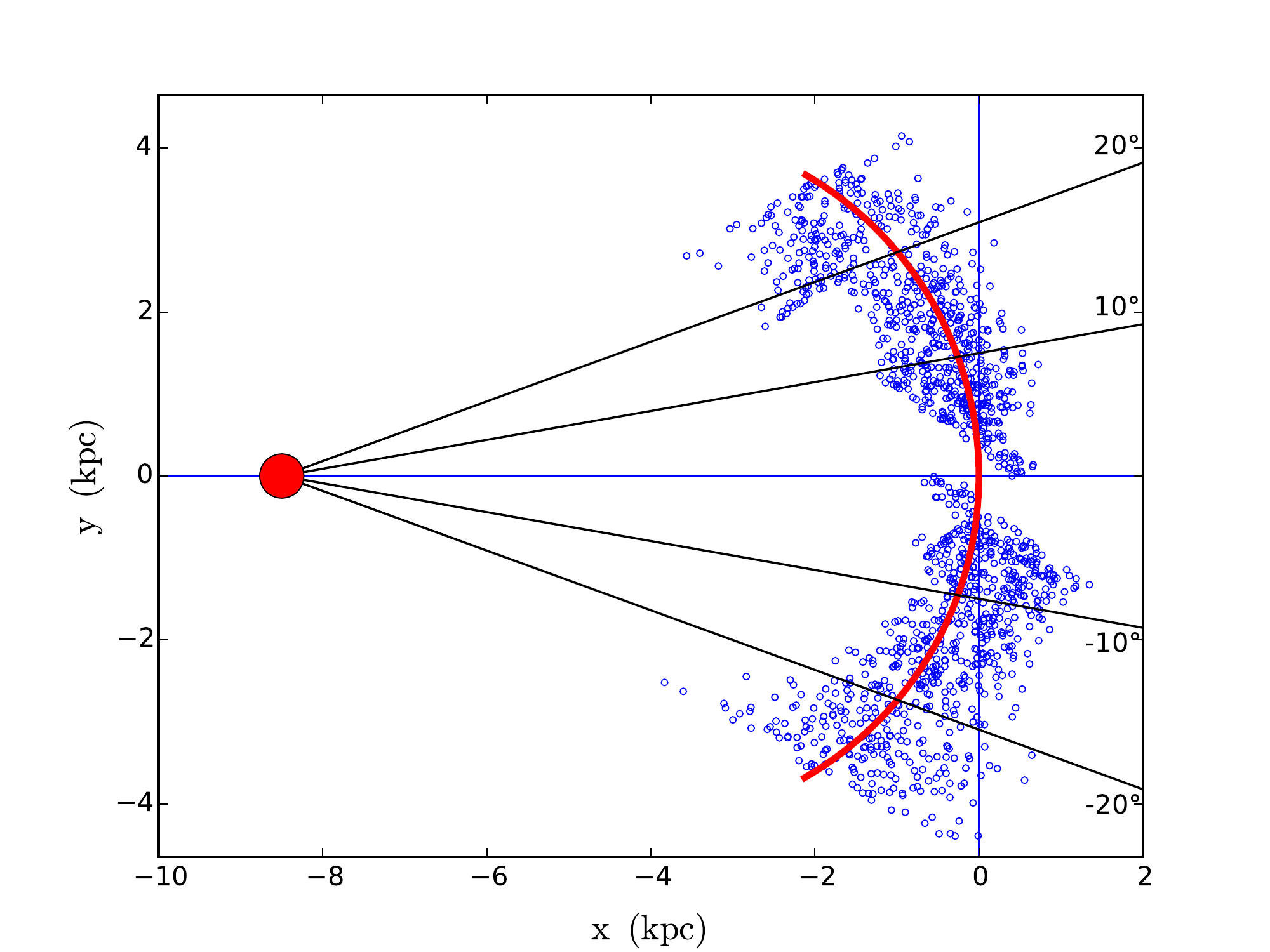}
\caption[]{Location of simulated \HI\ clouds selected according to the same 
criteria as were used in analysis of the the actual \HI\ data.  This gives an estimate of the area of the Galaxy 
sampled to produce Fig.~\ref{fig:image} and Fig.~\ref{fig:HI_z}.  The location of the Sun is marked with a red circle and lines are drawn 
at various longitudes.  The structure between longitudes 10\arcdeg\  and 20\arcdeg\  results from 
streaming along the model bar, which moves particles to velocities much lower than $V_t$.  A similar effect does 
not occur in the South.
\label{fig:WS_xy}}
\end{figure}

Figure~\ref{fig:WS_xy} shows a face-on view of the 
results of the simulation, where points mark locations of 
clouds that would be included in the tangent-point analysis according to the selection criteria used for the \HI.
This is our estimate of the area of the Galaxy that  contributes to Fig.~\ref{fig:image}.  
The choice of a constant  $\Delta v_1 = 50$ \kms\ for $\ell \leq 5\arcdeg$
 limits the accuracy of this simulation at $R \leq 0.75$, 
but in any case our results are incomplete so close to the nucleus where strong 
non-circular motions make kinematic analysis difficult \citep[e.g.,][]{BurtonLiszt78,Binney91,Rodriguez-Fernandez08}.  
The feature in the distribution around $10\arcdeg \lesssim \ell \lesssim  20\arcdeg$ 
marks the area where streaming along the 
bar causes particles to have a significantly reduced $\VLSR$ and thus not appear at $V_t$.  No similar region 
exists in the South.   
There are no sharp discontinuities in the spatial coverage of our analysis 
that could artifically remove a real extended \HI\ layer.  As is evident in 
Figure 1, the absence of \HI\ above and below the Galactic plane is a large-scale
 phenomenon in the inner Galaxy.

%-------------------------------------------- 
%Bibliography
%--------------------------------------------
\bibliographystyle{apj} 
\bibliography{references} %~mcg/tex/references.bib, bibtex file 
\normalsize

%---------------------------------------------
% Figures and Tables
%---------------------------------------------
\clearpage

%\begin{deluxetable}{ccc}
%\centering
%\tablecaption{Annular averages of \HI\ surface density
%\label{tab:surf_dens}}
%\tablehead{
%\colhead{Longitude range \tablenotemark{1}} & \colhead{Radius range (kpc)} & \colhead{$\Sigma_{HI}~{\rm
%    (M_{\odot}~pc^{-2})}$} }
%\startdata
%$10.2\arcdeg< |\ell| < 11.9\arcdeg $ & $1.50< R < 1.75$ & $0.4$ \\
%$13.6\arcdeg< |\ell| < 15.3\arcdeg $ & $2.00< R < 2.25$ &  $0.6$ \\
%$17.1\arcdeg< |\ell| < 18.9\arcdeg $ & $2.50< R < 2.75$ & $1.6$ \\
%$20.7\arcdeg< |\ell| < 22.5\arcdeg $ &$3.00< R < 3.25$ & $3.0$ \\
%$24.3\arcdeg< |\ell| < 26.2\arcdeg $ &$3.50< R < 3.75$ & $3.4$ \\
%\enddata
%\tablenotetext{1}{Combined data from longitude Quadrants I and IV.}
%\end{deluxetable}

\begin{deluxetable}{cc}
\centering
\tablecaption{Annular averages of \HI\ surface density
\label{tab:surf_dens}}
\tablehead{
\colhead{Radius range (kpc)} & \colhead{$\Sigma_{HI}~{\rm
    (M_{\odot}~pc^{-2})}$} }
\startdata
$1.50< R < 1.75$ & $0.4$ \\
$2.00< R < 2.25$ &  $0.6$ \\
 $2.50< R < 2.75$ & $1.6$ \\
$3.00< R < 3.25$ & $3.0$ \\
$3.50< R < 3.75$ & $3.4$ \\
\enddata
\end{deluxetable}

\end{document}